\documentclass[aps,prd,eqsecnum,floats,floatfix,twocolumn]{revtex4}

\usepackage{graphicx}
\usepackage{bm}

\newcommand{\third}{{\scriptstyle\frac{1}{3}}}
\newcommand{\half}{{\scriptstyle\frac{1}{2}}}
\newcommand{\fourth}{{\scriptstyle\frac{1}{4}}}
\newcommand{\eighth}{{\scriptstyle\frac{1}{8}}}
\newcommand{\sixteenth}{{\scriptstyle\frac{1}{16}}}
\newcommand{\thirtysecond}{{\scriptstyle\frac{1}{32}}}
\newcommand{\eqref}[1]{(\ref{#1})}

\begin{document}

\title{Dynamical Gauge Conditions for the Einstein Evolution Equations}

\author{Lee Lindblom and Mark A. Scheel}

\affiliation{Theoretical Astrophysics 130-33, California Institute of
Technology, Pasadena, CA 91125}

\date{\today}

\begin{abstract}
The Einstein evolution equations have previously been written in a
number of symmetric hyperbolic forms when the gauge fields---the
densitized lapse and the shift---are taken to be fixed functions of
the coordinates.  Extended systems of evolution equations are
constructed here by adding the gauge degrees of freedom to the set of
dynamical fields, thus forming symmetric hyperbolic systems for the
combined evolution of the gravitational and the gauge fields.  The
associated characteristic speeds can be made causal (i.e. less than or
equal to the speed of light) by adjusting 14 free parameters in these
new systems.  And 21 additional free parameters are available, for
example to optimize the stability of numerical evolutions.  The gauge
evolution equations in these systems are generalizations of the
``$K$-driver'' and ``$\Gamma$-driver'' conditions that have been used
with some success in numerical black hole evolutions.
\end{abstract}


\maketitle

\section{Introduction}
\label{s:introduction}

The traditional 3+1 approach to the study of the Einstein evolution
equations assumes that spacetime is foliated by a one-parameter family
of spacelike surfaces.  The spacetime metric is usually decomposed
with respect to these $t=$ constant surfaces according to
\begin{equation}
ds^2 = - N^2 dt^2 + g_{ij}(dx^i + N^i dt)(dx^j + N^j dt),
\end{equation}
where $g_{ij}$ is the (positive definite) three-metric on the $t=$
constant surfaces, and $N$ and $N^i$ are called the lapse and shift
respectively~\cite{york79}.  (The $x^i$ represent spatial coordinates
on the $t=$ constant surfaces.) If $\partial_t$ is the tangent vector
along which evolutions will be generated, and $\vec n=\partial_\tau$
is the unit normal to the $t=$ constant surfaces, then the above
definitions imply that $\partial_t = N\vec n + N^i\partial_i$. Thus
the lapse $N$ measures the rate at which proper time $\tau$ advances
(as a function of $t$) along the unit normals, while the shift $N^i$
measures the velocity of points with fixed spatial coordinates with
respect to the unit normals.

The lapse $N$ and shift $N^i$ are therefore descriptions of how the
coordinates $\{t,x^i\}$ are laid out on the spacetime manifold, and so
in this sense they represent coordinate or ``gauge'' degrees of
freedom.  The lapse and shift are not determined by the Einstein
equations, and may be chosen quite freely.  For example the Einstein
evolution equations have been written in a variety of symmetric
hyperbolic forms in which the (densitized) lapse and shift can be
specified as arbitrary functions of the coordinates
$\{t,x^i\}$~\cite{frittelli_reula94,
choquet_york95,friedrich96,frittelli_reula96,Bona1997,
anderson_etal98,anderson_york99,Frittelli1999,Alcubierre1999,Hern1999,
Kidder2001,Lindblom2002}.

Since the lapse and shift are not determined by the Einstein
equations, we have the opportunity and the responsibility to specify
them in some other manner.  We may use this freedom in a variety of
ways.  For example we could use it to simplify the representation of
the spacetime geometry (as is often done in spacetimes with
symmetries)~\cite{smarr77,petrich85,choptuik2003}, to simplify the
form of the evolution equations~\cite{fischer_marsden79}, to avoid
singularities (physical and
coordinate)~\cite{smarryork78,york79,Alcubierre2003}, or to attempt to
control the stability of numerical
evolutions~\cite{Alcubierre2001A,Alcubierre2002,Yo2002,Shoemaker2002}.

In this paper we develop systems of evolution equations that include
the lapse and the shift as dynamical fields.  These equations together
with appropriate versions of the Einstein evolution equations form
symmetric hyperbolic systems for the combined gravitational and gauge
fields.  Unified hyperbolic systems of equations for the evolution of
the gravitational and the gauge fields have been proposed before.  The
earliest of these uses harmonic gauge conditions that reduce the
evolution equations to a very simple
form~\cite{fischer_marsden79,Alvi2002}, but this gauge has not found
widespread use in numerical simulations of black hole spacetimes.
Dynamical gauge conditions have also been proposed that convert
well-known elliptic gauge conditions into equations that are
hyperbolic when the other dynamical fields are considered
fixed~\cite{bona96,Alcubierre2001A,Alcubierre2002}, but these
equations have never been fully integrated with the rest of the
Einstein evolution equations to form a unified hyperbolic system.
Strongly-hyperbolic~\cite{bona_masso95b} and more recently
symmetric-hyperbolic~\cite{Sarbach2002} formulations that include
rather general evolution equations for the lapse (but which still keep
the shift fixed) have also been proposed.  Here we propose a new
symmetric-hyperbolic system that includes dynamical equations for the
lapse and the shift.  Our equations are natural generalizations of the
``$K$-driver'' and the ``$\Gamma$-driver'' equations that have been
used with some success in evolving black hole
spacetimes~\cite{Alcubierre2001A,Alcubierre2002,Yo2002}.

In Sec.~\ref{s:gaugeequations} we review the properties of these gauge
evolution equations and in Sec.~\ref{s:evolution} combine them with
the Einstein evolution equations to form a single unified system.  In
Sec.~\ref{s:symmetrizer} we show that a 16-parameter family of these
combined (gauge and Einstein) evolution equations is symmetric
hyperbolic.  In Sec.~\ref{s:speeds} we find analytical expressions for
the characteristic speeds of these new systems.  These expressions
depend on 14 of the 16 free parameters.  We also demonstrate with
specific examples that all of these characteristic speeds can be made
causal ({\it i.e.,\/} less than or equal to the speed of light) by
making suitable choices for the 14 parameters.  Finally in
Sec.~\ref{s:kinematical} we extend the evolution equations by performing
a general kinematical transformation on the dynamical fields.  This
transformation depends on 19 additional free parameters which leave
the characteristic speeds and the hyperbolicity conditions unchanged.

\section{Dynamical Gauge Conditions}
\label{s:gaugeequations}

Our aim is to find equations for the gauge fields that allow the
spacetime coordinates to adapt dynamically to the structure of the
evolving spacetime.  In particular we would like the gauge fields to
select coordinates in which all the dynamical fields become time
independent whenever the spacetime itself evolves into an equilibrium
stationary state.  For computational efficiency and ease of
formulating appropriate boundary conditions, we prefer to find
hyperbolic rather than elliptic equations for the gauge fields. We
also prefer hyperbolic equations rather than equations of
indeterminant type because they have a well posed initial value
problem.

The desire to improve the stability and accuracy of numerical
evolutions of Einstein's equations has for many years provided the
motivation to find intelligent choices for the gauge
fields~\cite{smarryork78,york79}.  Perhaps the most widely studied
gauge condition of this type is the use of maximal-slice foliations
for the $t=$~constant surfaces in the 3+1 decomposition.  Maximal
slices are defined by the condition that the divergence of the normal
vector vanishes.  Maximal slices tend to avoid strong focusing
singularities, and they allow longer numerical evolutions than do
simpler choices such as $N=1$.  The mathematical condition that a
slice be maximal is equivalent to the condition that the trace of the
extrinsic curvature of the slice vanishes: $0 = K \equiv
g^{ij}K_{ij}$.  The time evolution of $K$ is determined by the
standard 3+1 ADM expression
\begin{equation}
\partial_t K - N^i\nabla_i K = -\nabla^i\nabla_i N + NK_{ij}K^{ij},
\end{equation}
where $\nabla_i$ is the covariant derivative compatible with $g_{ij}$.
Thus the choice of evolving along a foliation of maximal slices, each
with $K=0$, is enforced by imposing an elliptic equation on the lapse
$N$ .  This condition for the lapse is easily generalized to
conditions whose effect is to freeze $K$ to its value on an initial
surface: $0=\partial_t K$.  These ``$K$-freezing'' conditions also
result in elliptic equations for the lapse on each time slice:
\begin{equation}
\label{eq:BasicKFreezing}
0=-\partial_t K = \nabla^i\nabla_i N - NK_{ij}K^{ij}- N^i\nabla_i K.
\end{equation}

The $K$-freezing conditions have been used numerically with some
success~\cite{Alcubierre2001}.  One disadvantage is that they require
the solution of an elliptic equation at each time step. This is
usually more computationally expensive than solving hyperbolic
equations, and for the case of excised black
holes~\cite{seidel_suen92,bbhprl98a,Alcubierre2001,Brandt2000,
Alcubierre2001A,Yo2002,Scheel2002,Shoemaker2003} it requires
appropriate boundary conditions~\cite{Cook2002} to be imposed on the
excision surfaces.  For these reasons, alternatives to
Eq.~\eqref{eq:BasicKFreezing} have been studied as well.  One
possibility is to convert the elliptic equation for the lapse into a
hyperbolic equation, by adding suitable time derivative terms.  Thus
one might take
\begin{equation}
\partial_t^2 N + \kappa N \partial_t N = -\mu N^2 \partial_t K,
\label{e:lapsewave}
\end{equation}
as a gauge condition~\footnote{The factors of the lapse $N$ that
appear in this and subsequent equations were chosen to make these wave
equations more covariant.  Thus for example by using the coefficient
$\mu N^2$ instead of $\mu$ in Eq.~\eqref{e:lapsewave} the
characteristic speeds relative to a surface normal observer are
$\pm\sqrt{\mu}$ instead of $\pm\sqrt{\mu}/N$.}.  The second time
derivative term $\partial_t^2 N$ converts the elliptic equation for
$N$ into a hyperbolic equation with characteristic speeds
$\pm\sqrt{\mu}$, while the first-order term $\kappa N\partial_t N$
provides dissipation that tends to suppress $\partial_tN$.  Gauge
conditions of this type have been called ``$K$-driver''
conditions~\cite{Balakrishna1996} and have been used with some success
in the numerical evolution of black hole
spacetimes~\cite{Alcubierre2001A,Alcubierre2002,Shoemaker2002}.  A
large family of different $K$-driver conditions can be constructed
from Eq.~\eqref{e:lapsewave} by adding terms that leave the hyperbolic
structure of this equation intact. Here we will use as our starting
point one of these $K$-driver equations that admits an exact first
time integral. Thus we adopt a first-order $K$-driver condition which
can be thought of as the first-integral of an equation like
Eq.~\eqref{e:lapsewave}:
\begin{equation}
0=\partial_t N - N^i\partial_iN + \kappa N^2 + \mu N^2 (K - K_0).
\label{e:simpleKdriver}
\end{equation}
Here $K_0$ is the arbitrarily prescribed value of $K$ on some
$t=$~constant surface.  Lapse functions that solve this equation will
also satisfy a damped wave equation that is analogous to
Eq.~\eqref{e:lapsewave}.  Thus our expectation is that (if and) when a
spacetime evolves into a time-independent state, this choice of lapse
will drive the evolution toward a slicing in which the trace of the
extrinsic curvature $K$ takes the time-independent value $K_0$.

Next we turn our attention to finding appropriate conditions for the
shift $N^i$.  The idea is to use our freedom in the shift to select
spatial coordinates in which the evolution of the spatial metric
$\partial_t g_{ij}$ approaches zero whenever the spacetime itself
evolves toward a stationary state.  The time derivative of the
spatial metric is given by the usual 3+1 ADM expression,
\begin{equation}
\partial_t g_{ij} = \nabla_i N_j + \nabla_j N_i - 2 N K_{ij}\equiv\Sigma_{ij}.
\end{equation}
York~\cite{york79} showed that the integral of the square of
$\Sigma_{ij}+\bar\lambda g_{ij}g^{kl}\Sigma_{kl}$ over a $t=$~constant
surface is minimized whenever its divergence vanishes:
\begin{eqnarray}
0&=&\nabla_j\left(\Sigma^{ji}+\bar\lambda g^{ji}\Sigma\right)
\label{e:mindistortion}\\
& =&
\nabla_j\bigl(\nabla^jN^i+\nabla^iN^j+\bar\lambda g^{ji}\nabla_kN^k\bigr)
\nonumber\\&&
- 2 \nabla_j\left[N (K^{ji}+\bar\lambda g^{ji}K)\right].
\label{e:mindistortion1}
\end{eqnarray}
This is an elliptic equation for $N^i$ whenever
$\bar\lambda>-2$~\cite{Holst2002}.  Such a condition selects shift vectors
that minimize the time derivative of the spatial metric (or more
accurately the time derivative of the densitized metric $g^{\bar\lambda}
g_{ij}$), and includes the well-studied minimal distortion shift
condition (the case $\bar\lambda = -\third$)~\cite{york79}.

It would be straightforward to convert the shift conditions of
Eq.~\eqref{e:mindistortion1} to hyperbolic equations by adding
appropriate time derivative terms, in analogy with the derivation of the
$K$-driver equation for the lapse.  However, we choose instead to
follow a slightly different path.  Motivated by the work of
Alcubierre, {\it et.~al\/}~\cite{Alcubierre2001,Alcubierre2002} we consider the
quantity
\begin{equation}
\label{eq:defineGamma}
\tilde\Gamma^i \equiv \tilde g^{kl}\tilde\Gamma^i{}_{kl},
\end{equation}
where $\tilde g_{ij}$ is the conformal metric $\tilde g_{ij} \equiv
g^{\lambda} g_{ij}$, $g=\det g_{ij}$, and $\tilde\Gamma^i{}_{kl}$ is
the connection compatable with $\tilde g_{ij}$.  The quantity $\tilde
\Gamma^i$ agrees with the dynamical field used in the
BSSN~\cite{shibata95,baumgarte99} formulation of the Einstein
equations when $\lambda=-\third$.  It follows
from~(\ref{eq:defineGamma}) that
\begin{equation}
g^\lambda \tilde \Gamma^i = - g^{-(1+\lambda)/2}\partial_j
\bigl[g^{(1+\lambda)/2}g^{ji}\bigr],
\end{equation}
and
\begin{eqnarray}
\partial_t\bigl(g^\lambda\tilde\Gamma^i\bigr)&=&
\partial_j\left[\Sigma^{ji}-\half(1+\lambda)g^{ji}\Sigma\right]
\nonumber\\&&+\half(1+\lambda)\bigl[\Sigma^{ij}\partial_j\log g 
+\Sigma\partial_jg^{ji}\bigr].\qquad
\end{eqnarray}
Thus the ``$\Gamma$-freezing'' condition
$\partial_t(g^\lambda\tilde\Gamma^i)=0$ imposes an elliptic equation
on the shift (for $\lambda < 3$ in this case).  This
$\Gamma$-freezing differential equation has the same principal part as
the generalized minimum distortion condition,
Eq.~\eqref{e:mindistortion}.  Following Alcubierre, {\it
et.~al\/}~\cite{Alcubierre2002} we convert this elliptic shift condition
into a hyperbolic equation by adding appropriate time derivative
terms, {\it e.g.\/}
\begin{equation}
\partial_t^2 N^i + \kappa N \partial_t N^i 
= \mu N^2 \partial_t\bigl(g^\lambda \tilde \Gamma^i\bigr).
\label{e:shiftwave}
\end{equation}
As was the case for the lapse equation, it is possible to construct a
large family of hyperbolic $\Gamma$-freezing conditions by adding
non-principal terms to Eq.~\eqref{e:shiftwave}.  By adding suitable
non-principal terms we can construct members of this family that admit
exact first integrals. So we adopt as our ``$\Gamma$-driver''
condition one of these exact first integrals:
\begin{equation}
0=\partial_t N^i - N^j\partial_j N^i + \kappa N N^i  -\mu N^2
\bigl(g^\lambda\tilde
\Gamma^i-g^\lambda_0\tilde \Gamma^i_0\bigr).\label{e:simpleGammadriver}
\end{equation}
Here the time-independent $g^\lambda_0\tilde \Gamma^i_0$ is the value
of $g^\lambda\tilde\Gamma^i$ on some particular time slice.  Our
expectation is that (if and) when a spacetime evolves to a stationary
state, that the $\Gamma$-driver condition will cause the
spatial coordinates to evolve in a way that tends to minimize the
coordinate time derivatives of the spatial metric.

In summary then, we adopt the following $K$-driver and
$\Gamma$-driver conditions for the evolution of the lapse and shift:
\begin{eqnarray}
0&=& \partial_t N   - N^j \partial_j N   + \mu_L N^2 \bigl(K-K_0\bigr) 
\nonumber \\
                 &&\qquad+\kappa_L N^2-\epsilon_L N \partial_i N^i,
\label{e:Kdriver}\\
0&=&\partial_t N^i - N^j \partial_j N^i 
- \mu_S N^2 \bigl(g^\lambda \tilde\Gamma^i-g^\lambda_0 \tilde\Gamma^i_0\bigr)
		 \nonumber \\ 
	         &&\qquad+ \kappa_S N N^i
		  - \epsilon_S N g^{ij} \partial_j N.\qquad
\label{e:Gammadriver}
\end{eqnarray}
These conditions are just the $K$-driver and $\Gamma$-driver
conditions of Eqs.~\eqref{e:simpleKdriver} and
\eqref{e:simpleGammadriver} except for the addition of coupling terms
between the equations that are proportional to $\epsilon_L$ and
$\epsilon_S$.  These coupling terms will give us more flexibility
later in constructing a unified system of fully hyperbolic equations
for the evolution of all the gravitational and gauge fields.  For
maximum flexibility, at this stage we take the 7 parameters $\lambda$,
$\mu_L$, $\mu_S$, $\kappa_L$, $\kappa_S$, $\epsilon_L$ and
$\epsilon_S$ to be completely free and undetermined.

\section{Unified Evolution System}
\label{s:evolution}

The $K$-driver and $\Gamma$-driver Eqs.~\eqref{e:Kdriver} and
\eqref{e:Gammadriver} were each constructed to be first-order
hyperbolic equations.  However, these equations are manifestly
hyperbolic only when the other dynamical fields ({\it e.g.,\/}
$g_{ij}$, $K_{ij}$) are fixed, whereas the situation of interest to us
is when all fields evolve together.  So our aim now is to construct a
unified system of evolution equations for both the gauge and the
gravitational fields such that the entire system is symmetric
hyperbolic.

The first step is to examine the highest derivative coupling of the
(densitized) lapse and shift to the Einstein evolution equations.  We
use a general form of the equations written in the notation of
Kidder-Scheel-Teukolsky (KST)~\cite{Kidder2001}.  These are
first-order evolution equations for the spatial metric $g_{ij}$, the
extrinsic curvature $K_{ij}$ and the spatial derivatives of the metric
$D_{kij}=\half\partial_k g_{ij}$.  At this point we
need consider only the highest derivative (or
principal) parts of the equations:
\begin{eqnarray}
&&\!\!\!\!\!
\partial{}_t g{}_{ij}\simeq N^n\partial_n g_{ij} + 2 g_{n(i}\partial_{j)}N^n,
\label{e:kstggauge}\\
&&\!\!\!\!\!
\partial{}_t K{}_{ij}\simeq N^n\partial_n K_{ij}-N\partial_i\partial_j Q
-N\Bigl[(1+2\sigma)g^{cd}\delta^n{}_{(i}\delta^b{}_{j)}
\nonumber\\
&&\qquad\quad-(1+\zeta)g{}^{nd}\delta{}^b{}_{(i}\delta{}^c{}_{j)} 
-(1-\zeta) g{}^{bc}\delta{}^n{}_{(i}\delta{}^d{}_{j)} 
\nonumber\\&&\qquad\quad 
+g{}^{nb}\delta{}^c{}_i\delta{}^d{}_j 
+2\gamma g{}^{n[b}g{}^{d]c}g{}_{ij}\Bigr]
\partial{}_n D{}_{bcd},\label{e:kstkgauge}\\
&&\!\!\!\!\!
\partial{}_t D{}_{kij} \simeq N^n\partial{}_n D{}_{kij}
+g_{a(i}\partial_{j)}\partial_k N^a-N\Bigl[
\delta{}^n{}_k\delta{}^b{}_i\delta{}^c{}_j 
\nonumber\\
&&\qquad\quad-\half\eta g{}^{nb}g{}_{k(i}\delta{}^c{}_{j)}
-\half\chi g{}^{nb}g{}_{ij}\delta{}^c{}_{k}
+ \half\eta g{}^{bc}g{}_{k(i}\delta{}^n{}_{j)}\nonumber\\
&&\qquad\quad
+ \half\chi g{}^{bc}g{}_{ij}\delta{}^n{}_{k}\Bigr]\partial{}_n K{}_{bc}
,
\qquad\label{e:kstdgauge}
\end{eqnarray}
where $\simeq$ denotes equality of the principal part of the equation,
and $Q=\log\left(N/g^\sigma\right)$ is the densitized lapse.  The
parameter $\sigma$ that appears in these equations is part of the
definition of the densitized lapse $Q$, while $\gamma$, $\eta$,
$\chi$, and $\zeta$ were introduced by adding multiples of the
constraints to the evolution equations (see KST~\cite{Kidder2001}).

The Einstein evolution Eqs.~\eqref{e:kstkgauge} and
\eqref{e:kstdgauge} couple to the second spatial derivatives of the
densitized lapse and shift.  Thus in order to construct a first-order
unified system, we need to promote the spatial derivatives of the
gauge fields to the status of independent dynamical fields; so let
\begin{eqnarray}
T_i &=& \partial_i Q,\\
M_k{}^i &=& N^{-1}\partial_k N^i.
\end{eqnarray}
Using these definitions we express the gauge evolution
equations~\eqref{e:Kdriver} and \eqref{e:Gammadriver} in terms of
these new fields.  Furthermore, we obtain evolution equations for
$T_i$ and $M_k{}^i$ by taking spatial gradients of
Eqs.~\eqref{e:Kdriver} and \eqref{e:Gammadriver}. The principal parts
of the resulting equations are then given by~\footnote{Note that we
can change the principal parts of Eqs.~\eqref{e:qppdef} and
\eqref{e:nppdef} by adding terms proportional to
$N^k(\partial_kQ-T_k)$ and
$N^k(\partial_kN^i-N M_k{}^i)$. However, doing this for the shift
equation spoils linear degeneracy, making the system genuinely
nonlinear.  We chose to keep the system linearly degenerate since this
condition is known to prevent the formation of shocks in similar (but
not identical) hyperbolic systems~\cite{liu1979}.},
\begin{eqnarray}
\label{e:qppdef}
\partial_t Q       &\simeq& 0,\\
\label{e:nppdef}
\partial_t N^i     &\simeq& 0,\\
\label{e:tppdef}
\partial_t T_i     &\simeq& N^k\partial_k T_i + N(2\sigma-\mu_L)\partial_i K
		        \nonumber \\ 
			&+& N(\epsilon_L-2 \sigma) \partial_i M_j{}^j,\\
\label{e:mppdef}
\partial_t M_j{}^i &\simeq& N^k\partial_k M_j{}^i
	                + 2 N \mu_S g^{im}g^{kl}\partial_j D_{klm}
			\nonumber \\ 
	                &&+   N [2\epsilon_S\sigma-\mu_S(1+\lambda)]
				 g^{im}g^{kl}\partial_j D_{mkl}
			\nonumber \\ 
			&&+ \epsilon_S N g^{ik} \partial_j T_k.
\end{eqnarray}
In deriving the last two equations we made use of the constraints
\begin{eqnarray}
\label{e:Tconstraint}
{\cal C}_{ij} &\equiv&     2\partial{}_{[i}T{}_{j]}=0,\\
\label{e:Mconstraint}
{\cal C}_{nk}{}^i &\equiv& 2N^{-1}\partial{}_{[n}\bigl(N M_{k]}{}^i\bigr)=0,
\end{eqnarray}
in order to write all the terms involving $N^k$ as advection
terms~\footnote{We note that the 3-index constraint ${\cal
C}_{nk}{}^i$ used in this paper is not the same as the 3-index constraint
${\cal C}_{kij}$ introduced by KST~\cite{Kidder2001}.}.

The system of Eqs.~\eqref{e:kstggauge}--\eqref{e:kstdgauge} and
\eqref{e:qppdef}--\eqref{e:mppdef} constitutes a unified system of
first-order evolution equations for the full set of dynamical fields
$\{g_{ij},$ $K_{ij},$ $D_{kij},$ $Q,$ $N^i,$ $T_i,$ $M_k{}^i\}$ as
desired.  However, this system is not unique.  We are free to add
multiples of the various constraints to these equations, thus
producing other systems whose constraint-satisfying solutions are
identical.  Motivated by the fact that the addition of such constraint
terms improves the mathematical character of the Einstein evolution
equations~\cite{Kidder2001},
we now add additional multiples of the constraints to our
unified system of equations.  In particular we modify
Eqs.~\eqref{e:kstdgauge}, \eqref{e:tppdef},
and~\eqref{e:mppdef} as follows:
\begin{eqnarray}
\partial_t T_i     &=& \ldots + \half \psi_1 N {\cal C}_i 
	                      + \half \psi_2 N {\cal C}_{ki}{}^k,\\
\partial_t M_j{}^i &=& \ldots + \half \psi_3 N g^{ik}{\cal C}_{jk}
                              + \half \psi_4 N g^{ia}g^{bc}{\cal C}_{ajbc}
\nonumber\\&&\hspace{1.4em}
                              + \half \psi_5 N g^{ib}g^{ca}{\cal C}_{ajbc}
                              + \half \psi_6 N g^{ia}g^{bc}{\cal C}_{abcj}
\nonumber\\&&\hspace{1.4em}
                              + \half \psi_7 N \delta^i_j {\cal C},\\
\partial_t D_{kij} &=& \ldots + \half \psi_8 N g_{a(i}{\cal C}_{j)k}{}^a
                              + \half \psi_9 N g_{ij}{\cal C}_{ka}{}^a
\nonumber\\&&\hspace{1.4em}
                              + \half \psi_{10} N g_{k(i} {\cal C}_{j)a}{}^a.
\end{eqnarray}
Here the $\ldots$ denote the terms in the unmodified equations.  The
new terms (each proportional to a new constant $\psi_A$) include
multiples of the new constraints, ${\cal C}_{ij}$ and ${\cal
C}_{nk}{}^i$ of Eqs.~\eqref{e:Tconstraint} and~\eqref{e:Mconstraint},
as well as multiples of the standard Hamiltonian and momentum
constraints ${\cal C}$ and ${\cal C}_i$, and the constraint ${\cal
C}_{klij}$ from the fixed-gauge Einstein evolution system.  These
latter constraints are defined by~\footnote{We note that the
constraints are to be thought of as functions of $Q$, $N^i$, $g_{ij}$,
$K_{ij}$, $D_{kij}$, $T_i$, $M_k{}^i$ and the first derivatives
$\partial_k K_{ij}$, $\partial_l D_{kij}$, $\partial_k T_i$, and
$\partial_l M_k{}^i$.  The explicit expressions for the constraints in
terms of these quantities are given in Appendix~\ref{s:appendixa}.}
\begin{eqnarray}
{\cal C}       &=& \half\bigl[{}^{(3)}R-K _{ij} K ^{ij}+K^2\bigr],
\label{e:ch} \\
{\cal C}_i     &=& \nabla _j K ^j{}_i - \nabla _i K,\\
{\cal C}_{klij}&=& 2\partial_{[k}D_{l]ij}\label{e:ct}.
\end{eqnarray}
Adding constraints in this way is essential for obtaining a hyperbolic
system of evolution equations. Note that the constraints added
here are in addition to the constraints already included in
Eqs.~\eqref{e:kstkgauge} and \eqref{e:kstdgauge}.

The full unified system of evolution equations, including these new
constraint terms, can now be written as follows (showing here only the
principal parts):
\begin{eqnarray}
\partial_t g_{ij}  &\simeq& N^k\partial_k g_{ij},\label{e:newgeqnc}\\
\partial_t Q       &\simeq& 0,\label{e:newqeqnc}\\
\partial_t N^i     &\simeq& 0,\label{e:newnieqnc}\\
\partial_t T_i     &\simeq& N^k\partial_k T_i + N(2\sigma-\mu_L)\partial_i K
		        \nonumber \\ 
			&&+ N(\epsilon_L-2\sigma) \partial_i M_j{}^j
		        \nonumber \\ 
			&& +\psi_1 N \partial_{[k} K_{i]}{}^{k}
                           +\psi_2 N \partial_{[k}M_{i]}{}^k,
			    \label{e:newteqnc}\\
\partial_t M_j{}^i &\simeq& N^k\partial_k M_j{}^i
	                + 2 N \mu_S g^{im}g^{kl}\partial_j D_{klm}
			\nonumber \\ 
	                &&+   N [2\epsilon_S\sigma-\mu_S(1+\lambda)]
				 g^{im}g^{kl}\partial_j D_{mkl}
			\nonumber \\ 
			&&+ \epsilon_S N g^{ik} \partial_j T_k
			+\psi_{3}N g^{ik}\partial_{[j}T_{k]}\nonumber\\
&&+N\Bigl[\psi_4 g^{i[n}\delta^{a]}_j g^{bc}
+\psi_5 g^{i(b}g^{c)[n}\delta^{a]}_j
\nonumber\\&&
+\psi_{6} g^{i[n}g^{a](b}\delta^{c)}_j
+\psi_{7} g^{a[b}g^{n]c}\delta^i_j
\Bigr]\partial_n D_{abc},\qquad
\label{e:newmeqnc}\\
\partial{}_t D{}_{kij} &\simeq& N^n\partial{}_n D{}_{kij}
+N\Bigl[g{}_{b(i}\delta^n_{j)}\delta^a_k
+\psi_{8} g{}_{b(i}\delta^{[n}_{j)}\delta^{a]}_k
\nonumber\\&&
+\psi_{9} g_{ij}\delta^{[n}_k\delta^{a]}_b
+\psi_{10} g_{k(i}\delta^{[n}_{j)}\delta^{a]}_b
\Bigr]\partial_{n}M_a{}^b\nonumber\\
&&-N\Bigl[
\delta{}^n{}_k\delta{}^b{}_i\delta{}^c{}_j 
-\half\eta g{}^{nb}g{}_{k(i}\delta{}^c{}_{j)}
-\half\chi g{}^{nb}g{}_{ij}\delta{}^c{}_{k}\nonumber\\
&&
+ \half\eta g{}^{bc}g{}_{k(i}\delta{}^n{}_{j)}
+ \half\chi g{}^{bc}g{}_{ij}\delta{}^n{}_{k}\Bigr]\partial{}_n K{}_{bc}.
\label{e:newdeqnc}\\
\partial{}_t K{}_{ij}&\simeq& N^n\partial_n K_{ij}
-N\partial_{(i} T_{j)}-N\Bigl[
 (1+2\sigma) g{}^{cd}\delta{}^n{}_{(i}\delta{}^b{}_{j)}\nonumber\\
&&-(1+\zeta) g{}^{nd}\delta{}^b{}_{(i}\delta{}^c{}_{j)} 
-(1-\zeta) g{}^{bc}\delta{}^n{}_{(i}\delta{}^d{}_{j)}\nonumber\\ 
&&+g{}^{nb}\delta{}^c{}_i\delta{}^d{}_j 
+2\gamma g{}^{n[b}g{}^{d]c}g{}_{ij}\Bigr]
\partial{}_n D{}_{bcd}.\label{e:newkeqnc}
\end{eqnarray}
These equations constitute a first-order system of evolution equations
for the dynamical fields $\{g_{ij},$ $K_{ij},$ $D_{kij},$ $Q,$ $T_i,$
$ N^i,$ $M_k{}^i\}$.  This system depends on 22 freely specifiable
parameters: 20 of these parameters affect the principal parts of the
equations $\{\sigma,$ $\gamma,$ $\eta,$ $\chi,$ $\zeta,$ $ \psi_1,$
$\ldots,$ $\psi_{10},$ $\lambda,$ $\mu_L,$ $\mu_S,$ $ \epsilon_L,$
$\epsilon_S\}$, while 2 additional parameters $\{\kappa_L,$
$\kappa_S\}$ are dissipation terms in the gauge equations that do not
affect the principal parts.  We will constrain some of these
parameters in the following section to ensure that the system of
equations is symmetric hyperbolic.  The remaining parameters will be
freely specifiable and available for other purposes, such as
simplifying the resulting equations or optimizing the stability of
numerical spacetime evolutions.

These evolution equations for the dynamical fields $\{g_{ij},$
$K_{ij},$ $D_{kij},$ $Q,$ $T_i,$ $ N^i,$ $M_k{}^i\}$ also imply
evolution equations for the various constraints of the system.  In
Appendix~\ref{s:appendixa} we derive these constraint evolution
equations, and show that if the constraints are exactly satisfied
initially then they will continue to be satisfied as the system
evolves.

\section{Symmetric Hyperbolicity}
\label{s:symmetrizer}

The unified system of evolution
equations~(\ref{e:newgeqnc}--\ref{e:newkeqnc}) derived in
Sec.~\ref{s:evolution} can be written in the form
\begin{equation}
\partial_t u^\alpha + A^{k\,\alpha}{}_\beta\partial_k u^\beta\simeq 0,
\label{e:firstordersystem}
\end{equation}
where $u^\alpha$ is the collection of dynamical fields:
$u^\alpha=\{g_{ij},$ $K_{ij},$ $D_{kij},$ $Q,$ $T_i,$ $N^i,$ $M_k{}^i\}$.
A first-order system such as this is called symmetric hyperbolic if
there exists a symmetric positive-definite ``symmetrizer''
$S_{\alpha\beta}$ on the space of dynamical fields such that
$A^k_{\alpha\beta}\equiv S_{\alpha\mu}A^{k\mu}{}_\beta$
is symmetric for all $k$: $A^k_{\alpha\beta}=A^k_{\beta\alpha}$.
Symmetric hyperbolic systems~\cite{courant-hilbert} have well-posed
initial value problems, real characteristic speeds, complete sets of
characteristic eigenvectors, and other nice mathematical properties
such as the existence of associated canonical energy norms.

We now explore the conditions under which the unified evolution
equations of Sec.~\ref{s:evolution} are in fact symmetric hyperbolic.
We assume that the symmetrizer $S_{\alpha\beta}$ can be written as a
function of the metric $g_{ij}$ and various constant parameters.  In
particular we consider the following general symmetrizer which we
express as a quadratic form~\footnote{We have considered the most
general symmetrizer that can be constructed from the spatial metric
and constant parameters.  The symmetrizer given in Eq.~\eqref{e:sab}
however lacks a number of cross terms between the metric variables
$\{g_{ij},Q,N^i\}$ and the derivative variables
$\{K_{ij},D_{kij},T_i,M_k{}^i\}$.  These cross terms must vanish as a
consequence of the symmetry conditions, and so for simplicity we omit
them from Eq.~\eqref{e:sab}.},
\begin{eqnarray}
&&dS^2\equiv S_{\alpha\beta}du^\alpha du^\beta \nonumber\\
&&= 
A_1\, dG^2 +A_2\, g{}^{ik}g{}^{jl}d\tilde g_{ij} \, d\tilde g_{kl}
+A_3 dQ^2 
\nonumber\\ &&\,\,\,
+ A_4 g_{ij}dN^idN^j+ B_1\, dK^2 
+B_2\, g{}^{ik}g{}^{jl}d\tilde K_{ij} \, d\tilde K_{kl}
\nonumber\\&&\,\,\,
+C_1 \, g{}^{kl}g{}^{ia}g{}^{jb} d\tilde D {}_{(kij)} \, d\tilde D^{}_{(lab)}
\nonumber\\ &&\,\,\,
+C_2 \, g{}^{kl}g{}^{ia}g{}^{jb} \bigl[d\tilde D {}_{kij}
		                      -d\tilde D {}_{(kij)}\bigr] 
				 \bigl[d\tilde D {}^{}_{lab}
				      -d\tilde D {}^{}_{(lab)}\bigr]
\nonumber\\ &&\,\,\,
+C_3  \, g^{ij} dD^{1}_{i} \, dD^{1}_{j}
+C_4  \, g^{ij} dD^{2}_{i} \, dD^{2}_{j}
+2C_5 \, g^{ij} dD^{1}_{i}dD^{2}_{j} 
\nonumber\\ &&\,\,\,
+E_1  g^{ij} dT_i \, dT_j 
+2D_1 g^{ij} dT_i \, dD^{1}_{j} 
+2D_2 g^{ij} dT_i \, dD^{2}_{j}
\nonumber\\ &&\,\,\,
+ E_2 dM^2 
+ {\scriptscriptstyle {1\over 2}} 
E_3 \left[g_{ij} g^{kl}+\delta_i^l \delta_j^k\right] 
                 d\tilde M_k{}^i \, d\tilde M_l{}^j
\nonumber\\ &&\,\,\,
+ {\scriptscriptstyle {1\over 2}} 
E_4 \left[g_{ij} g^{kl}-\delta_i^l \delta_j^k\right] 
                 d\tilde M_k{}^i \, d\tilde M_l{}^j
\nonumber\\ &&\,\,\,
+ 2 D_3 dM dK + 2 D_4 g^{ik} \delta^j_l d\tilde K_{ij} \, d\tilde M_k{}^l.
\label{e:sab} 
\end{eqnarray}
Here $dG$, $dK$ and $dM$ are the traces of $dg _{ij}$, $dK {}_{ij}$
and $dM_k{}^i$
respectively, and $d\tilde g {}_{ij}$ , $d\tilde K {}_{ij}$
and $d\tilde M_k{}^i$ are their
trace-free parts.  The two
traces of $dD {}_{kij}$ are defined by
\begin{eqnarray}
dD {}^{1}_{i}&\equiv&g{}^{jk}dD {}_{ijk}\\ 
dD {}^{2}_{i}&\equiv&g{}^{jk}d D {}_{kij},
\end{eqnarray}
and its trace-free part, $d\tilde D{}_{kij}$, is
\begin{eqnarray}
d\tilde D {}_{kij} \equiv dD {}_{kij}&+& {\scriptstyle \frac{1}{5}}
\bigl[dD {}^{1}_{(i}g {}_{j)k}-2\,dD {}^{1}_{k} g {}_{ij}\nonumber\\
      &&\quad+ dD {}^{2}_{k} g {}_{ij}-3\, dD {}^{2}_{(i}g {}_{j)k}\bigr].
\end{eqnarray}

The quadratic form~\eqref{e:sab} is positive definite iff the
parameters corresponding to diagonal symmetrizer elements $\{A_1,$
$A_2,$ $A_3,$ $A_4,$ $B_1,$ $B_2,$ $C_1,$ $C_2,$ $C_3,$ $C_4,$ $E_1,$
$E_2,$ $E_3,$ $E_4\}$ are positive, and certain inequalities are
satisfied by the parameters corresponding to the off-diagonal
symmetrizer elements $\{C_5,$ $D_1,$ $D_2,$ $D_3,$ $D_4\}$.  Some of
these off-diagonal inequalities are simple, i.e. $D_3^2 < E_2 B_1$,
and $D_4^2 < B_2 E_3$.  But the inequalities involving the other 
off-diagonal parameters $\{C_5,D_1,D_2\}$ are less transparent.  The
needed condition is that the $3\times3$ matrix
\begin{equation}
\left(\begin{array}{ccc}
 C_3 & C_5 & D_1\\
 C_5 & C_4 & D_2\\
 D_1 & D_2 & E_1\\
\end{array}\right)\label{e:matrix}
\end{equation}
is positive definite.  The most straightforward way to enforce this
condition is to use the fact that a matrix is positive definite
iff it admits a Cholesky decomposition~\cite{numrec_c}. 
By writing the Cholesky decomposition of Eq.~\eqref{e:matrix}
in terms of new parameters $F_A$, we obtain
\begin{eqnarray}
C_3&=&F_1^2+F_2^2+F_3^2,\\
C_4&=&F_4^2+F_5^2,\\
C_5&=&F_2F_4+F_3F_5,\\
D_1&=&F_3F_6,\\
D_2&=&F_5F_6,\\
E_1&=&F_6^2.
\end{eqnarray}
Given these expressions for $\{C_3,$ $C_4,$ $C_5,$ $D_1,$ $D_2,$
$E_1\}$, the matrix in Eq.~\eqref{e:matrix} is positive definite for
arbitrary $F_A$, so long as $F_1\neq 0$, $F_4\neq 0$, and $F_6\neq 0$.

It is straightforward (but tedious) now to evaluate the conditions on
the various parameters needed to guarantee that the matrices
$A{}^k{}_{\alpha\beta}=S{}_{\alpha\mu}A{}^{k\mu}_{\beta}$ are
symmetric in $\alpha$ and $\beta$ for all $k$.  After lengthy
algebraic manipulations we find that the following conditions are
necessary and sufficient to guarantee that the $A{}^k{}_{\alpha\beta}$ are
symmetric:
\begin{eqnarray}
B_2&=&(C_1+2C_2)/3,\label{e:b2}\\
\zeta&=&-C_1(E_3+D_4)/[E_3B_2-D_4^2],\\
\sigma&=&\!\Bigl\{3(3D_3+2D_4)\bigl[-3(2C_1+5C_3+5C_4+10C_5)E_1
\nonumber\\
&&+5(D_1+D_2)(3D_1+3D_2+3D_3+2D_4)\bigr]\nonumber\\
&&-(3E_2+2E_3)\bigl[9E_1(2C_1+5C_3+5C_4+10C_5)\nonumber\\
&&+5(D_1+D_2)\bigl(9(B_1-D_1-D_2)+2C_1+4C_2\bigr)\bigr]
\Bigr\}\nonumber\\
&&/\bigl[-10E_1(3E_2+2E_3)(9B_1+2C_1+4C_2)\nonumber\\
&&+30E_1(3D_3+2D_4)^2\bigr],\\
\mu_S&=&\bigl(16C_1+5C_2+15C_4\bigr)/30(E_3+E_4),\\
\psi_8&=&2(C_1-C_2+3D_4)/3C_2,\\
\epsilon_S&=& \Bigl\{5E_1(3E_2+2E_3)(1-\lambda)\mu_S-6E_1C_1
\nonumber\\
&&-15(C_3+C_4+2C_5)E_1-10E_1(3D_3+2D_4)\sigma\nonumber\\
&&+5(D_1+D_2)\bigl[3(D_1+D_2+D_3)+2D_4\bigr]\Bigr\}\nonumber\\
&&/5(3E_2+2E_3)(D_1+D_2-2\sigma E_1),\\
\mu_L&=&2\sigma+\bigl[3B_1+2B_2-3(D_1+D_2)\nonumber\\
&&-(3D_3+2D_4)\epsilon_S\bigr]/3E_1,\\
\epsilon_L&=&2\sigma-\bigl[3(D_1+D_2+D_3)+2D_4\nonumber\\
&&-(3E_2+2E_3)\epsilon_S\bigr]/3E_1,\\
\psi_3&=& -(1+E_3/E_4)\epsilon_S+(D_2+D_4)/E_4,\\
\psi_4&=&\bigl\{2C_1-5C_2-15\bigl[C_5+(1+2\sigma)D_4\bigr]\nonumber\\
&&-15(E_3+E_4)\bigl[(1+\lambda)\mu_S-2\sigma \epsilon_S\bigr]\bigr\}/15E_4,\\
\psi_5&=&(C_1+2C_2+3D_4)/3E_4+(C_1-\zeta D_4)/E_3,\\
\psi_6&=&(C_1+2C_2+3D_4)/3E_4-(C_1-\zeta D_4)/E_3,\\
\gamma&=&\Bigl\{\bigl[3C_1-15(C_3+C_5)+30\sigma D_1\bigr](D_3+E_2)
\nonumber\\
&&+5(6B_1+B_2)E_2-5(6D_3+D_4+3\epsilon_L D_1)D_3\nonumber\\
&&-15\mu_L D_1E_2+10(D_4E_2-D_3E_3)(1+\lambda)\mu_S\nonumber\\
&&-10\sigma(3D_3+2D_4)D_3+20\sigma\epsilon_S(D_3E_3-D_4E_2)\nonumber\\
&&+10\sigma(3B_1+2B_2)E_2\Bigr\}/45(D_3^2-B_1E_2),\\
\chi&=&\Bigl\{\bigl[2(C_3+2C_5)E_1-2(D_1+2D_2)D_1\bigr]\nonumber\\
&&\times\bigl[30\mu_S D_4 E_1
+(16C_1+5C_2+15C_4)E_1\nonumber\\
&&+15(B_2-D_2-\epsilon_S D_4)D_2\bigr]\nonumber\\
&&+\bigl[(2C_4+C_5)E_1-(D_1+2D_2)D_2\bigr]\nonumber\\
&&\times\bigl[30(D_2+\epsilon_SD_4)D_1
-2(8C_1-5C_2+15C_5)E_1\nonumber\\
&&+15D_4E_1(2\mu_S+2\lambda\mu_S-\psi_5+\psi_6-4\sigma\epsilon_S)\nonumber\\
&&-30B_2(D_1-2\sigma E_1+\zeta E_1)\bigr]\Bigr\}
/75E_1\bigl[C_3D_2^2\nonumber\\
&&+C_5^2E_1-2C_5D_1D_2
+(D_1^2-C_3E_1)C_4\bigr],\\
\eta&=&\bigl[15(2-\chi)D_2^2+15(2\epsilon_SD_4-3\chi D_1)D_2\nonumber\\
&&+8C_1E_1+10C_2E_1+15(\psi_5-\psi_6-4\mu_S)D_4E_1\nonumber\\
&&+15(\chi-2)C_4E_1-30(D_2+E_1-\zeta E_1)B_2\nonumber\\
&&+45\chi C_5E_1\bigr]/15\bigl[(D_1+2D_2)D_2-(2C_4+C_5)E_1\bigr],\nonumber\\
&&\\
\psi_2&=&\Bigl\{-16C_1(C_4D_1+2C_3D_2-2C_5D_1-C_5D_2)\nonumber\\
&&-5C_2\bigl[C_4D_1+C_5(D_1-D_2)-C_3D_2\bigr](4+3\psi_8)\nonumber\\
&&+30(C_3C_4-C_5^2)\bigl[D_4+(E_4-E_3)\epsilon_S+E_4\psi_3\bigr]
\nonumber\\
&&+60(C_3D_2-C_5D_1)(E_3-E_4)\mu_S\nonumber\\
&&-(C_4D_1-C_5D_2)\Upsilon
\Bigr\}/30\bigl[E_1(C_5^2-C_3C_4)
\nonumber\\
&&
+C_3D_2^2+C_4D_1^2
-2C_5D_1D_2\bigr],\\
\psi_7&=& \bigl[2C_1-5C_2+15C_4+30C_5-60D_3-90\gamma D_3\nonumber\\
&&-10D_4-10(E_3+6E_2-3E_4)\mu_S\nonumber\\
&&+30(\epsilon_L-2\sigma)D_2\bigr]/45E_2,\\
\psi_1&=&\bigl[-2B_2-(3\chi+\eta)D_1\nonumber\\
&&+(2-\chi-2\eta)D_2+2D_4\epsilon_S\bigr]/E_1,\\
\psi_{10}&=&\Bigl\{(3C_3+C_5)\bigl[60(E_4-E_3)\mu_S-5C_2(4+3\psi_8)
\nonumber\\
&&+32C_1+30D_2\psi_2-15C_4\psi_8+30C_5(2+\psi_8)\bigr]\nonumber\\
&&-(C_4+3C_5)\bigl[16C_1+5C_2(4+3\psi_8)+30D_1\psi_2\nonumber\\
&&-15C_5\psi_8+30C_3(2+\psi_8)+\Upsilon
\bigr]\Bigr\}
\nonumber\\&&
/150(C_3C_4-C_5^2),\\
\psi_9&=&\bigl[32C_1-5(4+3\psi_8)C_2-15(\psi_8+4\psi_{10})C_4\nonumber\\
&&+30(2+\psi_8-\psi_{10})C_5+60(E_4-E_3)\mu_S\nonumber\\
&&+30\psi_2D_2\bigr]/30(C_4+3C_5),\label{e:p9}
\end{eqnarray}
where $\Upsilon$ is given by
\begin{eqnarray}
\Upsilon&=& 30(1+\lambda)(E_3-E_4)\mu_S
+30D_4(2\sigma-\zeta)\nonumber\\
&&-15E_4(2\psi_4+\psi_5+\psi_6-4\sigma\epsilon_S)\nonumber\\
&&+15E_3(\psi_6-\psi_5-4\sigma\epsilon_S).\label{e:upsilon}
\end{eqnarray}

These conditions determine the 20 parameters $\{B_2,$ $\psi_1,$
$\ldots,$ $\psi_{10},$ $\sigma,$ $\gamma,$ $\eta,$ $\chi,$ $\zeta,$
$\mu_L,$ $\mu_S,$ $\epsilon_L,$ $\epsilon_S\}$ in terms of the 15
parameters $\{\lambda,$ $B_1,$ $C_1,$ $C_2,$ $C_3,$ $C_4,$ $C_5,$
$D_1,$ $D_2,$ $D_3,$ $D_4,$ $E_1,$ $E_2,$ $E_3,$ $E_4\}$~\footnote{We
note that Eqs.~\eqref{e:b2}--\eqref{e:p9} are ordered so that each of
the parameters $\{B_2,$ $\psi_1,$ $\ldots,$ $\psi_{10},$ $\sigma,$
$\gamma,$ $\eta,$ $\chi,$ $\zeta,$ $\mu_L,$ $\mu_S,$ $\epsilon_L,$
$\epsilon_S\}$ appears on the right-hand side of only those equations
that follow the one in which it appears on the left.}.  Writing the
conditions for symmetric hyperbolicity like this is a particularly
convenient way to parameterize these evolution systems.  The
parameters $\{\lambda,$ $B_1,$ $C_1,$ $C_2,$ $C_3,$ $C_4,$ $C_5,$
$D_1,$ $D_2,$ $D_3,$ $D_4,$ $E_1,$ $E_2,$ $E_3,$ $E_4\}$ can be chosen
freely except for the simple inequalities needed to guarantee the
positivity of $S_{\alpha\beta}$.  We note that the evolution system is
invariant under an overall scaling of the symmetrizer.  Thus without
loss of generality we will set $C_1=1$, so there are really only 14
freely specifiable parameters that affect the evolution equations.  We
also point out the following nice feature of this way of
parameterizing these equations: By using the symmetrization conditions
in Eqs.~\eqref{e:b2}--\eqref{e:upsilon} to determine the parameters
that actually appear in the evolution equations $\{\psi_1,$ $\ldots,$
$\psi_{10},$ $\sigma,$ $\gamma,$ $\eta,$ $\chi,$ $\zeta,$ $\mu_L,$
$\mu_S,$ $\epsilon_L,$ $\epsilon_S\}$ we are guaranteed to have a
system that has only real characteristic speeds, a complete set of
eigenvectors, etc.  This same parameterization technique has been used
by Frittelli and Reula~\cite{Frittelli1999}, and can also be used to
provide a more convenient and complete characterization of the
symmetric hyperbolic subset of the original fixed-gauge KST equations.
We summarize this approach to the KST equations in the Appendix.
Finally we note that the four symmetrizer parameters
$\{A_1,A_2,A_3,A_4\}$ do not enter any of the symmetry conditions.  So
while these parameters can be chosen quite freely (ensuring only that
they are positive) they do not seem to play any important role in
determining the dynamics of the system.

\section{Characteristic Speeds}
\label{s:speeds}

The evolution equations for the full system of fields---including the
gauge fields---have been put in a first-order form in
Sec.~\ref{s:evolution}.  The characteristic speeds in the direction
$\xi_k$ are defined as the eigenvalues of the matrix $\xi_k
A^{k\,\alpha}{}_\beta$ that appears in
Eq.~\eqref{e:firstordersystem}. The unit one-form $\xi_k$ specifies
the direction of propagation.  The characteristic speeds associated
with the fields $\{g_{ij},Q,N^i\}$ are very simple.  In the frame of
the hypersurface-normal observers, the characteristic speed associated
with the propagation of $g_{ij}$ is $v=0$, while the speed associated
with the propagation of the gauge fields $\{Q,N^i\}$ is $v=-\xi_k
N^k/N$~\footnote{The characteristic speeds associated with these
fields are actually quite arbitrary.  For example we could change them
to arbitrary multiples of $-\xi_kN^k/N$ (including zero) simply by
adding the terms $V^k(\partial_kQ-T_k)$ and $V^k(\partial_k N^i -
N M_k{}^i)$ to Eqs.~\eqref{e:newqeqnc} and \eqref{e:newnieqnc}
with $V^k$ an arbitrary multiple of $N^k$.  Terms of this type do not
change the symmetric hyperbolicity of the system, but they can affect
its linear degeneracy (and hence the ability of the system to form
shocks). The vector $V^k$ that appears in these extra terms could also
be specified as an arbitrary function of the coordinates.  In this
case these characteristic speeds become $-\xi_k(N^k-V^k)/N$ which can
be adjusted to arbitrary values with appropriate choice of $V^k$ while
leaving the system linearly degenerate.  These speeds can all be made
less than the speed of light in this way if that is desired.}.

In order to evaluate the characteristic speeds associated with the
other dynamical fields of this system it is convenient to transform to
an irreducable representation of the space of fields.  In this basis
the matrix $\xi_k A^{k\,\alpha}{}_\beta$ becomes block diagonal and
hence its eigenvalues become much easier to evaluate.  The irreducable
representation of the remaining dynamical fields
$\{T_i,$ $M_k{}^i,$ $K_{ij},$ $D_{kij}\}$ consists of projecting them onto the
scalars $\{T_i\xi^i,$ $M_k{}^i\xi^k\xi_i,$ $M_k{}^iP^k{}_i,$
$K_{ij}\xi^i\xi^j,$ $K_{ij}P^{ij},$ $D_{kij}\xi^k\xi^i\xi^j,$
$D_{kij}\xi^kP^{ij},$ $D_{kij}P^{ki}\xi^j\}$ (where
$P_{ij}=g_{ij}-\xi_i\xi_j$ is the projection tensor onto the two-space
orthogonal to $\xi_i$), the transverse vectors $\{T_iP^i{}_j,$
$M_k{}^i\xi^kP_i{}^j,$ $M_k{}^i\xi_iP^{kj},$ $K_{ik}\xi^iP^{kj},$
$D_{kil}\xi^k\xi^iP^{lj},$ $D_{kil}P^{kj}\xi^i\xi^l,$
$D_{kil}P^{k(i}P^{l)j},$ $D_{kil}P^{kj}P^{il}\},$ the symmetric
transverse traceless tensors $\{M_{kl}P^{kl}{}_{ij},$
$K_{kl}P^{kl}{}_{ij},$ $D_{mkl}\xi^mP^{kl}{}_{ij},$
$D_{mkl}\xi^lP^{mk}{}_{ij}\}$ (where
$P^{kl}{}_{ij}=P^k{}_iP^l{}_j-\half P^{kl}P_{ij}$), the antisymmetric
transverse tensors $\{M_k{}^l P^k{}_{[i}P_{j]l},$
$K_{kl}P^k{}_{[i}P^l{}_{j]}\},$ and finally the transverse traceless
part of $D_{kij}$.
 
The scalar parts of the dynamical fields form an eight dimensional
subspace, and this $8\times 8$ block of $\xi_k A^{k\,\alpha}{}_\beta$
decouples from the others.  This block depends on the dynamical KST
parameters and the other parameters introduced in
Sec.~\ref{s:evolution} that describe the dynamics of the gauge fields.
We find that the eight characteristic speeds (relative to the normals
of the hypersurface) can be represented as,
\begin{eqnarray}
v^2_{S1\pm}&=& A_{S1}\pm B_{S1},\\
v^2_{S2\pm}&=& A_{S2}\pm B_{S2},
\end{eqnarray}
where
\begin{eqnarray}
A_{S1}&=&\half \bigl[\mu_L+(1-\lambda)\mu_S +\epsilon_S\epsilon_L\bigr],\\
B^2_{S1}&=&A^2_{S1}+(1-\lambda)(\epsilon_L-\mu_L)\mu_S,\\
A_{S2}&=&\fourth\bigl[(1+2\gamma)(2+2\chi-\eta)-\eta\zeta\bigr]
\nonumber\\
&&+\sixteenth(2+\psi_8+2\psi_{10})(\psi_5-\psi_6)\nonumber\\
&&+\eighth\psi_7(2+3\psi_8-4\psi_9+2\psi_{10}),\qquad\quad\\
B^2_{S2}&=&A^2_{S2}-\eighth
\bigl[(1+\chi)(2+\psi_8+2\psi_{10})+\eta(\psi_8-2\psi_9)\bigr]\nonumber\\
&&\qquad\quad\times
\bigl[(1+2\gamma)(\psi_5-\psi_6)-2\zeta\psi_7\bigr].
\end{eqnarray} 

The transverse vector parts of the dynamical fields constitute two
identical eight dimensional subspaces.  The eight characteristic speeds
for each of these blocks are,
\begin{eqnarray}
v^2_{V0}&=&0,\\
v^2_{V1}&=& \mu_S,\\
v^2_{V2\pm}&=& A_{V2}\pm B_{V2},
\end{eqnarray}
where
\begin{eqnarray}
A_{V2}&=&-\eighth(\psi_1+\psi_2\psi_3)
+\eighth\psi_4(2+\psi_8-3\psi_9-\psi_{10})
\nonumber\\
&&+\eighth\psi_{5}(2+\psi_8-\psi_9)-\sixteenth\psi_6(2\psi_9+3\psi_{10})
\nonumber\\
&&+\sixteenth\eta(1-3\zeta-4\sigma)-\eighth\chi(1+6\sigma),\\
B^2_{V2}&=&A^2_{V2}+\chi \Lambda_\chi+\eta \Lambda_\eta + \Lambda,\\
\Lambda_\chi&=&\sixteenth
(2+\psi_8)\bigl[(1+3\zeta)\psi_4+(1+4\sigma+\zeta)\psi_5
\nonumber\\
&&-(2\sigma-\zeta)\psi_6\bigr]-\thirtysecond\psi_{10}\bigl[(5-9\zeta)\psi_4
\nonumber\\&&+
(1-4\sigma-3\zeta)\psi_5+(4+14\sigma-3\zeta)\psi_6\bigr]\nonumber\\
&&-\sixteenth\psi_2\bigl[(1+6\sigma)\psi_3+3\psi_4+\psi_5+\psi_6\bigr],\\
\Lambda_\eta&=&\thirtysecond
\psi_2\bigl[(1-4\sigma-3\zeta)\psi_3-2\psi_4-3\psi_6\bigr]
\nonumber\\
&&+\thirtysecond(2+\psi_8)
\bigl[3(\zeta-2\sigma)\psi_6-(1-4\sigma-3\zeta)\psi_5
\nonumber\\
&&\quad+(5\zeta-1)\psi_4
\bigr]
+\thirtysecond\psi_9\bigl[(1-4\sigma-3\zeta)\psi_5
\nonumber\\
&&\quad+(5-9\zeta)\psi_4+(4+14\sigma-3\zeta)\psi_6
\bigr]
,\\
\Lambda&=&\thirtysecond
\psi_1\psi_{10}\bigl[(1-4\sigma-3\zeta)\psi_3-2\psi_4-3\psi_6\bigr]
\nonumber\\
&&+\sixteenth\psi_1(2+\psi_8)\bigl[\psi_5+(2\sigma-\zeta)\psi_3+\psi_4\bigr]
\nonumber\\
&&-\sixteenth\psi_1\psi_9
\bigl[\psi_5+(1+6\sigma)\psi_3+3\psi_4+\psi_6\bigr].\end{eqnarray}
We note that the parameter $\mu_S$ introduced in Eq.~\eqref{e:Gammadriver} 
represents one of the characteristic speeds of this system, as expected.

The characteristic speeds of the two identical 4-dimensional spaces of
symmetric transverse traceless second rank tensors are:
\begin{eqnarray}
v^2_{(TT)1}&=& 1,\\
v^2_{(TT)2}&=& \eighth(\psi_5-\psi_6)(2+\psi_8).
\end{eqnarray}
The characteristic speeds of the 2-dimensional space of antisymmetric
transverse-traceless second-rank tensors are:
\begin{eqnarray}
v^2_{[TT]}&=&\eighth(\psi_5+\psi_6)(2+\psi_8).
\end{eqnarray}
And finally the subspace consisting of the transverse traceless part
of $D_{kij}$ has only one characteristic speed, and this vanishes.

These expressions determine the characteristic speeds in terms of the
parameters $\{\psi_1, \psi_2, ...\}$ that define the form of the evolution
equations.  The speeds can also be re-expressed in terms of the
symmetrizer parameters through Eqs.~\eqref{e:b2}--\eqref{e:upsilon}.
The characteristic speeds are therefore functions of the 14
parameters $\{B_1,$ $C_2,$ $C_3,$ $C_4,$ $C_5,$ $D_1,$ $D_2,$ $D_3,$
$D_4,$ $E_1,$ $E_2,$ $E_3,$ $E_4,$ $\lambda\}$ that can be specified
(almost) freely as discussed in Sec.~\ref{s:symmetrizer}.

Although most of the characteristic speeds depend on the parameters
that define the system of evolution equations, several of the speeds
are independent of them.  For instance, twelve eigenvectors have
characteristic speed zero, and four eigenvectors have characteristic
speed $\pm 1$ (the speed of light in our units); the former correspond
to gauge-dependent fields, while the latter must be the incoming and
outgoing fields corresponding to the two physical gravitational
degrees of freedom.  The remaining speeds, the adjustable ones, must
correspond to various gauge-dependent and therefore basically
unphysical characteristic fields.

In the past it has been considered most
natural~\cite{choquet_york97a,anderson_etal98,Kidder2001} to set any
adjustable speeds in the Einstein evolution equations to one (the
speed of light) or zero with respect to the $t=$ constant surface
normals.  Our experience, however, is that the instabilities limiting
evolutions of black hole spacetimes (with excision) often occur in
outgoing characteristic fields that propagate at the speed of light
just outside the event horizon~\cite{Lindblom2002}.  Excitations in
such fields remain in the computational domain for long periods of
time and therefore have the opportunity to grow large. It therefore
might be better to set the adjustable characteristic speeds to values
significantly less than the speed of light for evolutions of black
hole spacetimes.

We have not been able to show that the adjustable characteristic
speeds can be set to arbitrary values by adjusting the available
parameters, and in fact it appears likely that this is not
possible. In particular we have not been able to find parameter values
that make all of these speeds equal to unity or zero.  However, we
have shown that parameter values can be chosen to make all of the
adjustable characteristic speeds causal (i.e. less than or equal to
the speed of light)~\footnote{The characteristic speeds associated
with the fields $Q$ and $N^i$ are $-\xi_i N^i/N$ and cannot be
changed by adjusting the various parameters of the theory.  These
speeds will less than the speed of light except when the vector
$\partial_t$ becomes spacelike.  However as noted above
the equations can be easily modified to make these speeds causal
if that is desired.}. To provide a specific example, we
have have found parameter values that make the characteristic speeds
take the following ``simple'' values,
\begin{eqnarray}
0&=& v^2_{(TT)2}=v^2_{[TT]},\label{e:v2example0}\\
\fourth &=& v^2_{S1-}=v^2_{S2-}=v^2_{V2-},\\
\half &=& v^2_{S1+}=v^2_{S2+}=v^2_{V1}=v^2_{V2+},\label{e:v2example}
\end{eqnarray}
to any desired accuracy.  The approximate values of the symmetrizer
parameters needed to achieve these characteristic speeds are:
$B_1\!=\!7.17,\, C_1\!=\!1.00,\, C_2\!=\!2.57,\, C_3\!=\!8.68,\,
C_4\!=\!3.95,\, C_5\!=\!-3.81,\, D_1\!=\!5.36,\, D_2\!=\!4.86,\,
D_3\!=\!-10.78,\, D_4\!=\!-2.04,\, E_1\!=\!44.64,\, E_2\!=\!19.39,\,
E_3\!=\!3.65,\, E_4\!=\!2.22,\,\lambda\!=\!-0.33.$ These symmetrizer
parameters also determine the parameters that define the explicit form
of the evolution equations; for this example the latter parameters
have the following approximate values:
$\psi_1\!=\!0.13,\,\psi_2\!=\!-0.34,\,\psi_3\!=\!4.54,\,\psi_4\!=\!-0.92,\,
\psi_5\!=\!0.00,\,\psi_6\!=\!0.00,\,\psi_7\!=\!-0.90,\,\psi_8\!=\!-2.00,\,
\psi_9\!=\!-0.23,\,\psi_{10}\!=\!0.27,\,\gamma\!=\!-0.76,\,
\sigma\!=\!0.50,\,\zeta\!=\!-0.49,\,\eta\!=\!0.93,\,\chi\!=\!-0.43,\,
\mu_S\!=\!0.50,\,\mu_L\!=\!0.63,\,
\epsilon_S\!=\!-1.24,\,\epsilon_L\!=\!0.44.$ We see that all of the
characteristic speeds in this example are causal, and the various
parameters that determine the evolution equations are all of order
unity.  In another example, we explored the possibility of making all
of the characteristic speeds which appear in
Eq.~\eqref{e:v2example0}--\eqref{e:v2example} as small as possible.
We found that it was only possible to make the squares of all these
characteristic speeds smaller than about 0.29.  Thus it is relatively
easy to find examples of these evolution equations that appear to be
reasonable candidates for performing numerical evolutions of black
hole spacetimes.

\section{Kinematical Extension}
\label{s:kinematical}

Finally we note that the independent dynamical fields
$u^\alpha\equiv\{g_{ij},$ $K_{ij},$ $D_{kij},$ $Q,$ $T_i,$ $ N^i,$
$M_k{}^i\}$ can also be modified in these evolution equations.  It has
been shown~\cite{Kidder2001,Lindblom2002,Scheel2002} that seemingly
trivial changes in the choice of these dynamical fields can have
dramatic effects on the stability of numerical spacetime
evolutions. So following KST~\cite{Kidder2001} we introduce a set of
linear transformations on the dynamical fields.  In particular we take
a new set of fields $\hat u^\alpha$ defined by a transformation of the
form $\hat u^\alpha = T^\alpha{}_\beta u^\beta$, where
$T^\alpha{}_\beta$ depends only on various parameters and the spatial
metric $g_{ij}$.  The most general such transformation (which
preserves the fundamental metric fields $\{g_{ij},Q,N^i\}$)
is~\footnote{This is an extension (with a slight change in notation)
of the transformation introduced by KST~\cite{Kidder2001}.  The
relationship between the original KST notation and that used here is:
$\hat z_1 = \hat z$, $\hat k_1 = \hat k$, $\hat k_2 = \hat e$, $\hat
k_3 = \hat a$, $\hat k_4 = \hat b$, $\hat k_5 = \hat c$, and $\hat k_6
= \hat d$.}:
\begin{eqnarray}
\label{eq:kinematicKev}
\hat K_{ij} &=& K_{ij} + \hat z_1 g_{ij} g^{ab} K_{ab} 
+ \hat z_2 M_{(i}{}^ag_{j)a}\nonumber\\
&& + \hat z_3 g_{ij}M_a{}^a,\\
\label{eq:kinematicMev}
\hat M_k{}^i &=& \hat z_4 M_k{}^i + \hat z_5 \delta _k^i M_a{}^a
+\half\hat z_6 \left(\delta^a_k\delta^i_b-g^{ia}g_{kb}\right)M_a{}^b
\nonumber\\
&&+\hat z_7 K_{ka}g^{ai} + \hat z_8 \delta_k^i g^{ab}K_{ab},\\
\label{eq:kinematicDev}
\hat D_{kij} &=& \Bigl[\hat k_1 \delta^a_k\delta^b_i\delta^c_j
+\hat k_2 \delta^a_{(i}\delta^b_{j)}\delta^c_k
+\hat k_3 g_{ij}g^{bc}\delta^a_k\nonumber\\
&&+\hat k_4 g_{ij}g^{ab}\delta^c_k
+\hat k_5 g_{k(i}\delta^a_{j)}g^{bc}
+\hat k_6 g_{k(i}\delta^c_{j)}g^{ab}\Bigr]D_{abc}\nonumber\\
&&+\hat k_7 T_k g_{ij}+\hat k_8 g_{k(i}T_{j)},\\
\label{eq:kinematicTev}
\hat T_i &=& \hat k_9 T_i + \hat k_{10} g^{ab} D_{iab}
+\hat k_{11} g^{ab}D_{abi}.
\end{eqnarray}
This transformation is (generically) invertable, and is the identity
when $\hat z_4=\hat k_1=\hat k_9 =1$ and all other $\hat z_A$ and
$\hat k_A$ vanish.  Note that as in KST~\cite{Kidder2001}, it is
understood that when constructing evolution equations for the
transformed fields $\hat u^\alpha$, all (temporal and spatial)
derivatives of $g_{ij}$ that are introduced by differentiating
$T^\alpha{}_\beta$ are to be eliminated by substituting the definition
of $D_{kij}$ and the evolution equations for $g_{ij}$.

We also note that the kinematical transformations $\hat u^\alpha =
T^\alpha{}_\beta u^\beta$ described by
Eqs.~\eqref{eq:kinematicKev}--\eqref{eq:kinematicTev} do not change the
hyperbolicity of the system or the characteristic speeds. This is
because the characteristic matrix for the transformed system is $\xi_k
\hat{A}^{k\,\alpha}{}_\beta\equiv \xi_k T^\alpha{}_\mu
A^{k\,\mu}{}_\nu (T^{-1})^\nu{}_\beta$, which has the same eigenvalues
as $\xi_k A^{k\,\alpha}{}_\beta$, and the symmetrizer for the
transformed system is $\hat{S}_{\alpha\beta}\equiv S_{\mu\nu}
(T^{-1})^\mu{}_\alpha (T^{-1})^\nu{}_\beta$, which is symmetric and
positive definite iff $S_{\mu\nu}$ is symmetric and positive definite.

In summary, the unified system of evolution equations presented here
contains 41 free parameters when written in terms of the dynamical
fields $\hat u^\alpha$: the 22 parameters that entered
Eqs.~\eqref{e:newgeqnc}--\eqref{e:newkeqnc} as desribed above, plus 19
transformation parameters $\{\hat z_1, \ldots,\hat z_8,\hat
k_1,\ldots,\hat k_{11}\}$.  In Sec.~\ref{s:symmetrizer} we reduced the
number of free parameters by 6 to ensure that the system of equations
is symmetric hyperbolic.  The remaining 35 parameters are freely
specifiable and available for other purposes, such as simplifying the
resulting equations, fixing the characteristic speeds to desired
values, or optimizing the stability of numerical spacetime evolutions.

\acknowledgments

We thank Michael Holst, Markus Keel, Lawrence Kidder, Harald Pfeiffer,
and Manuel Tiglio for helpful conversations concerning this work.
This research was supported in part by NSF grant PHY-0099568 and NASA
grant NAG5-10707.

\appendix

\section{Constraint Evolution}
\label{s:appendixa} 

The dynamical system derived in Sec.~\ref{s:evolution} represents the
Einstein evolution equations only when certain constraints are
satisfied.  Here we derive the system of evolution equations that the
constraints themselves satisfy.  We begin by giving explicit
expressions for each of the constraints in terms of the dynamical
fields and their spatial derivatives:
\begin{eqnarray}
{\cal C} &=& \bigl(g^{ik}g^{jl}-g^{ij}g^{kl}\bigr)\partial_l D_{kij}
+g^{i[j}g^{a]b}K_{ij}K_{ab}\nonumber\\
&&-\half\bigl(g^{kc}g^{ij}g^{ab}+2g^{ka}g^{ib}g^{jc}+8g^{k[i}g^{a]j}g^{cb}
\nonumber\\
&&-3g^{kc}g^{ia}g^{jb}\bigr)D_{kij}D_{cab},\label{e:acham}\\
{\cal C}_i &=& 
\bigl(g^{aj}g^{bk}\delta^{c}_i+g^{cj}g^{ab}\delta^{k}_i
-2g^{ca}g^{bj}\delta^{k}_i\bigr)D_{cab}K_{jk}\nonumber\\
&&+2 g^{j[k}\delta^{l]}{}_i\partial_k K_{jl},\label{e:acmom}\\
{\cal E}_i &=& \partial_iQ - T_i,\label{e:aet}\\
{\cal E}_{k}{}^i &=& N^{-1}\partial_k N^i - M_k{}^i,\label{e:aem}\\
{\cal E}_{kij} &=& \partial_k g_{ij} - 2 D_{kijj},\label{e:aed}\\
{\cal C}_{ij} &=& 2\partial_{[i}T_{j]},\label{e:act}\\
{\cal C}_{nk}{}^i &=&2 \partial_{[n}M_{k]}{}^i  
+ 2M_{[k}{}^i\bigl(T_{n]} + 2\sigma D_{n]ab}g^{ab}\bigr),\label{e:acm}\\
{\cal C}_{klij} &=& 2 \partial_{[k}D_{l]ij}.\label{e:acd}
\end{eqnarray}
We note that these expressions do not contain any spatial derivatives
of the metric fields $Q$, $N^a$ or $g_{ij}$: these derivatives were
replaced by $T_i$, $M_k{}^i$ and $D_{kij}$ wherever they occured.

The collection of constraints,
\begin{eqnarray}
c^\alpha &=& \{{\cal C},{\cal C}_i,{\cal E}_i,{\cal E}_k{}^i,{\cal E}_{kij},
{\cal C}_{ij},{\cal C}_{nk}{}^i,{\cal C}_{klij}\},\label{e:calphadef}
\end{eqnarray}
is a function of the set of dynamical fields $\{g_{ij},$ $K_{ij},$ $D_{kij},$
$Q,$ $T_i,$ $ N^i,$ $M_k{}^i\}$, hence the evolution
of the constraints is determined
by the dynamical evolution equations~(\ref{e:newgeqnc})--(\ref{e:newkeqnc}).
A straightforward (but very lengthy) calculation
shows that the evolution of the constraints is determined by a
system of the form
\begin{eqnarray}
\partial_t c^\alpha + C^{k\alpha}{}_\beta(u)\partial_k c^\beta =
F^\alpha{}_\beta(u,\partial u)c^\beta.\label{e:ceq}
\end{eqnarray}
The coefficients $C^{k\alpha}{}_\beta$ that appear in this expression
depend on the metric fields $Q$, $N^i$, and $g_{ij}$ as well as the
various parameters that enter
Eqs.~(\ref{e:newgeqnc})--(\ref{e:newkeqnc}).  The coefficients
$F^\alpha{}_\beta$ depend on all of the dynamical fields and their
derivatives.  The complete expressions for the constraint evolution
equations are very lengthy, so here we
list only the principal parts of
the equations,
$\partial_tc^\alpha\simeq-C^{k\alpha}{}_\beta(u)\partial_k c^\beta$,
which can be written:
\begin{eqnarray}
\partial_t{\cal C} &\simeq& N^n\partial_n {\cal C}-\half N (2 + 2\chi -\eta)
g^{ij}\partial_i{\cal C}_j \nonumber\\
&&+\fourth N( 2 + 2\psi_{10} + 3 \psi_8 - 4\psi_9)
g^{ij}\partial_i {\cal C}_{jk}{}^k\!,\label{e:achamdot}\\
\partial_t{\cal C}_i &\simeq& -N(1+2\gamma)\partial_i{\cal C}
+\half N (1+\zeta) g^{jk}g^{ab}\partial_j{\cal C}_{aikb}\nonumber\\
&&-\half N g^{jk}\partial_j{\cal C}_{ki}
-\half N (1+2\sigma) g^{jk}g^{ab}\partial_j{\cal C}_{kiab}\nonumber\\
&&+N^n\partial_n {\cal C}_i
+\half N (1-\zeta) g^{jk}g^{ab}\partial_j{\cal C}_{kabi},\label{e:acmomdot}\\
\partial_t{\cal E}_i &\simeq& 0,\label{e:aetdot}\\
\partial_t{\cal E}_{k}{}^i &\simeq& 0,\label{e:aemdot}\\
\partial_t{\cal E}_{kij} &\simeq& 
N^n\partial_n {\cal E}_{kij},\label{e:aeddot}\\
\partial_t{\cal C}_{ij} &\simeq& N\psi_1 \partial_{[i} {\cal C}_{j]}
+\half N\psi_2 \partial_k{\cal C}_{ij}{}^k,\label{e:actdot}\\
\partial_t{\cal C}_{nk}{}^i &\simeq& N^j \partial_j{\cal C}_{nk}{}^i
+N\psi_7 \delta^i{}_{[k}\partial_{n]}{\cal C}
+ N\psi_3 g^{ij}\partial_{[n}{\cal C}_{k]j}\nonumber\\
&&-N\psi_4 g^{ij}g^{ab}\partial_{[n}{\cal C}_{k]jab}
-N\psi_5 g^{ij}g^{ab}\partial_{[n}{\cal C}_{k]abj}\nonumber\\
&&+N\psi_6 g^{ij}g^{ab}\partial_{[n}{\cal C}_{|jab|k]},\label{e:acmdot}\\
\partial_t{\cal C}_{lkij} &\simeq& N^a\partial_a{\cal C}_{lkij}
+N g_{a(i}\partial_{j)}{\cal C}_{lk}{}^a
+N \psi_9 g_{ij} \partial_{[l} {\cal C}_{k]a}{}^a
\nonumber\\ 
&&+N\chi g_{ij} \partial_{[l}{\cal C}_{k]}
- \half N \eta \bigl(g_{i[l}\partial_{k]}{\cal C}_{j}+
g_{j[l}\partial_{k]}{\cal C}_{i}\bigr)
\nonumber\\
&&-\half N \psi_{8} \bigl( g_{ai}\partial_{[l}{\cal C}_{k]j}{}^a+
g_{aj}\partial_{[l}{\cal C}_{k]i}{}^a\bigr)\nonumber\\
&&-\half N \psi_{10} \bigl( g_{i[l}\partial_{k]}{\cal C}_{ja}{}^a+
g_{j[l}\partial_{k]}{\cal C}_{ia}{}^a\bigr).\label{e:acddot}
\end{eqnarray}
We note that the principal parts of the constraint evolution equations
can be written in a number of different ways using various identities
that the constraints satisfy:
\begin{eqnarray}
\partial_{[i}{\cal E}_{j]}&=& \half {\cal C}_{ji},\\
\partial_{[k}{\cal E}_{j]}{}^i
&=& 
\half{\cal C}_{jk}{}^i
+N^{-1}{\cal E}_{[k}{}^i\partial_{j]}N
\nonumber \\ &&
+M_{[k}{}^i\left({\cal E}_{j]}
	           +\sigma{\cal E}_{j]ab}g^{ab}\right),\\
\partial_{[l}{\cal E}_{k]ij}&=& {\cal C}_{klij},\\
\partial_{[k}{\cal C}_{ij]}&=&0,\\
\partial_{[n}{\cal C}_{jk]}{}^i
&=&
 {\cal C}_{[kj}{}^i\left(T_{n]}+2\sigma D_{n]ab}g^{ab}\right)
\nonumber \\ &&
+M_{[n}{}^i\left({\cal C}_{jk]}+2\sigma{\cal C}_{jk]ab}g^{ab}\right)
\nonumber \\ &&
+4\sigma M_{[n}{}^i{\cal E}_{k}{}^{ab}D_{j]ab},\\
\partial_{[n}{\cal C}_{kl]ij}&=&0.
\end{eqnarray}
We have not yet explored how the character of the constraint evolution
equations (e.g. their hyperbolicity) is affected by changing their
principal parts using these identities.

We note that Eq.~\eqref{e:ceq} implies that the time derivative of the
constraints vanishes, $\partial_t c^\alpha=0$, whenever the
constraints are satisfied, $c^\alpha=0$, at some initial time.  This
guarantees that the constraints will be satisfied for all time in
analytic spacetimes.  And if these constraint evolution equations are
symmetric hyperbolic (as we expect, but have not yet verified) then
this will guarantee quite generally that the constraints remain
satisfied for all time if they are satisfied initially.

\section{The KST System}
\label{s:appendixb}
The fixed-gauge version of the Einstein evolution equations proposed
by KST can be shown to be strongly or even symmetric hyperbolic for
certain choices of the five ``dynamical'' free parameters $\{\sigma,$
$\zeta,$ $\gamma,$ $\eta,$ $\chi\}$ characterizing that
system~\cite{Kidder2001}.  For the case in which all of the adjustable
characteristic speeds are set equal to the speed of light, we have
previously~\cite{Lindblom2002} determined the regions of this
parameter space where the equations are symmetric hyperbolic, and the
regions where they are not.

Using the technique developed in Sec.~\ref{s:symmetrizer}, we can now
construct all of the symmetric hyperbolic fixed-gauge KST systems
explicitly, even for the general case in which the adjustable speeds
are left unspecified.  For the fixed-gauge KST evolution system,
Eqs.~\eqref{e:kstggauge}--\eqref{e:kstdgauge}, the five dynamical
parameters $\{\sigma,$ $\zeta,$ $\gamma,$ $\eta,$ $\chi\}$ can be
written in terms of the symmetrizer parameters $\{B_1,$ $C_1,$ $C_2,$
$C_3,$ $C_4,$ $C_5\}$.  These conditions, which are equivalent to the
conditions for symmetric hyperbolicity in these systems, are
\begin{eqnarray}
B_2&=& (C_1+2C_2)/3,\\ 
\zeta&=& -3C_1/(C_1+2C_2),\label{e:KSTzeta}\\ 
\gamma&=&-\frac{32C_1+10C_2+45(C_4+C_5+2B_1)}{135B_1},\\
\sigma&=&\frac{18C_1+45(C_3+C_4+2C_5)}{10(9B_1+2C_1+4C_2)},\,\,\,
\label{e:KSTsigma}\\
\eta&=&{\scriptscriptstyle \frac{6}{5}}+ B_2\bigl[5(3C_3+C_4+4C_5)
+20(C_4+3C_5)\sigma\nonumber\\
&&-3(9C_3+C_4+6C_5)\zeta\bigr]/25(C_3C_4-C_5^2),\\
\chi&=&-{\scriptscriptstyle \frac{2}{5}}-B_2\bigl[5(C_3+2C_4+3C_5)
+20(2C_4+C_5)\sigma \nonumber\\
&&-3(3C_3+2C_4+7C_5)\zeta\bigr]/25(C_3C_4-C_5^2).
\end{eqnarray}
Specifying the parameters $\{\sigma,$ $\zeta,$ $\gamma,$ $\eta,$
$\chi\}$ in this way guarantees that the characteristic speeds of
this system,
\begin{eqnarray}
v_1^2 &=& 2\sigma,\\
v_2^2 &=& \eighth\eta(1-3\zeta-4\sigma)-\fourth\chi(1+6\sigma),\\
v_3^2 &=& \half\bigl[(1+2\gamma)(2+2\chi-\eta)-\eta\zeta\bigr],
\end{eqnarray}
are real, that the characteristic eigenvectors of the system are
complete, etc.  We also note that the expression for the parameters
$\zeta$ and $\sigma $ in Eqs.~\eqref{e:KSTzeta} and \eqref{e:KSTsigma},
enforce the conditions that $\zeta$ and $\sigma$ be limited to the
ranges $-3<\zeta<0$ and $0<\sigma$ for the symmetric hyperbolic
fixed-gauge KST systems no matter what values the characteristic
speeds may have.


\end{document}